\documentclass[12pt,preprint]{emulateapj}

\shorttitle{RSGs as Cosmic Abundance Probes: star clusters in M83, and the MZR of nearby galaxies}
\shortauthors{Davies et al.}
\usepackage{graphicx}
\usepackage{natbib}
\usepackage{color}
\usepackage{array}

\def\hii{H\,{\sc ii}}
\def\hei{He\,{\sc i}}

\def\mgi{Mg\,{\sc i}}
\def\fei{Fe\,{\sc i}}
\def\sii{Si\,{\sc i}}
\def\tii{Ti\,{\sc i}}
\def\ga{\mathrel{\hbox{\rlap{\hbox{\lower4pt\hbox{$\sim$}}}\hbox{$>$}}}}
\def\la{\mathrel{\hbox{\rlap{\hbox{\lower4pt\hbox{$\sim$}}}\hbox{$<$}}}}

\def\msun{$M$\mbox{$_{\normalsize\odot}$}}
\def\zsun{$Z$\mbox{$_{\normalsize\odot}$}}

\def\lsun{$L$\mbox{$_{\normalsize\odot}$}}

\def\kms{\,km~s$^{-1}$}

\def\arcsec{$^{\prime \prime}$}

\def\teff{$T_{\rm eff}$}

\def\logg{$\log g$}

\def\um{$\mu$m}
\def\chisq{$\chi^{2}$}

\newcommand{\fig}[1]{Fig.\ \ref{#1}}

\newcommand{\temptext}[1]{  #1  }

\begin{document}
\title{Red Supergiants as Cosmic Abundance Probes: massive star clusters in M83, and the mass-metallicity relation of nearby galaxies} 
  \author{Ben Davies\altaffilmark{1},
  Rolf-Peter Kudritzki\altaffilmark{2},
  Carmela Lardo\altaffilmark{3,1}, 
  Maria Bergemann\altaffilmark{4}, Emma Beasor\altaffilmark{1}, 
  Bertrand Plez\altaffilmark{5},
  Chris Evans\altaffilmark{6}, Nate Bastian\altaffilmark{1}, Lee R.\ Patrick\altaffilmark{7,8} }

\affil{$^{1}$Astrophysics Research Institute, Liverpool John Moores 
University, Liverpool Science Park ic2, \\ 146 Brownlow Hill, Liverpool, L3 5RF, UK.}

\affil{$^{2}$Institute for Astronomy, University of Hawaii, 2680
Woodlawn Drive, Honolulu, HI, 96822, USA} 

\affil{$^{3}$Laboratoire d?astrophysique, Ecole Polytechnique F\'{e}ed\'{e}rale de Lausanne (EPFL), Observatoire, 1290, Versoix, Switzerland} 

\affil{$^{4}$Max-Planck Institute for Astronomy, 69117, Heidelberg, Germany} 

\affil{$^{5}$Laboratoire Univers et Particules de Montpellier,
  Universit\'{e} de Montpellier 2, CNRS, F-34095 Montpellier, France}
  
\affil{$^{6}$UK Astronomy Technology Centre, Royal Observatory
  Edinburgh, Blackford Hill, Edinburgh., EH9 3HJ, UK}

\affil{$^{7}$Instituto de Astrofísica de Canarias, E-38205 La Laguna, Tenerife, Spain;
Universidad de La Laguna, Dpto Astrofísica, E-38206 La Laguna, Tenerife, Spain}

\affil{$^{8}$Institute for Astronomy, Royal Observatory Edinburgh, Blackford Hill, Edinburgh.,
  EH9 3HJ, UK}

\begin{abstract}
We present an abundance analysis of seven super-star clusters in the disk of M83. The near-infrared spectra of these clusters are dominated by Red Supergiants, and the spectral similarity in the $J$-band of such stars at uniform metallicity means that the integrated light from the clusters may be analysed using the same tools as those applied to single stars. Using data from VLT/KMOS we estimate metallicities for each cluster in the sample. We find that the abundance gradient in the inner regions of M83 is flat, with a central metallicity of $[Z] = 0.21 \pm 0.11$ relative to a Solar value of \zsun=0.014, which is in excellent agreement with the results from an analysis of luminous hot stars in the same regions. Compiling this latest study with our other recent work, we construct a mass-metallicity relation for nearby galaxies based entirely on the analysis of RSGs. We find excellent agreement with the other stellar-based technique, that of blue supergiants, as well as with temperature-sensitive  (`auroral' or `direct') \hii-region studies. Of all the \hii-region strong-line calibrations, those which are empirically calibrated to direct-method studies (N2 and O3N2) provide the most consistent results.
\end{abstract}

\keywords{}


\section{Introduction} \label{sec:intro}

The knowledge of galaxy chemical composition is a crucial constrain for galactic evolution studies in the local universe and at larger redshift \citep[e.g.,][]{Schaye15}.  Specifically, the observed  relationship between a galaxy's mass in stars and its central metallicity \citep[the mass-metallicity relation, or {\it MZR}, e.g.][]{lequeux79,Tremonti04}, as well as the trend of metallicity with the galactocentric distance \citep[e.g.][]{Zaritsky94} offer fundamental insights into several physical processes; including clustering, merging,  galactic winds, star formation history, and initial mass function \citep[][]{Zahid14,Kudritzki15}.

Given this importance of precise metallicity estimates, it is crucial to understand the systematic errors present in such measurements. The vast majority of the metallicity determinations in star-forming galaxies comes from the analyses of strong \ion{H}{2}-region emission lines (the so-called {\em strong-line} methods). Such methods were developed as a work-around to the fact that the temperature-sensitive {\it auroral} lines are very weak and difficult to detect, especially at high metallicity where the gas cools more efficiently, meaning that the electron temperature $T_e$ cannot be independently constrained. Various techniques to measure \ion{H}{2}-region metallicities from the strong-lines only have been established, either from calibration against photoionization models \citep[e.g.][]{Tremonti04} or by bootstrapping to \ion{H}{2}-region observations where the $T_e$-sensitive lines are detectable \citep[e.g.][]{Pettini-Pagel04}. However, these strong-line techniques are potentially subject to large and poorly-understood systematic errors, which appear to become larger with increasing metallicity \citep[up to $\leq$0.7 dex depending on the adopted calibration, e.g.][]{Ercolano07,K-E08,Bresolin09}. Errors of this magnitude undermine the diagnostic power of abundance information obtained by such methods \citep[e.g.,][]{Kudritzki15}. 

Quantitative analysis studies of supergiant stars, the brightest stars in galaxies, have been recently performed in an attempt to overcome these calibration issues.  At optical wavelengths, low resolution observations of extragalactic blue supergiants (BSGs) have provided us with very accurate metallicities for galaxies beyond the Local Group \citep[e.g.][]{Kudritzki08,Kudritzki12,Kudritzki13,Kudritzki14, Kudritzki16,Hosek14,Bresolin16}, whilst at near-infrared (IR) wavelengths red supergiant stars (RSGs) represent powerful probes to study galaxy chemical composition. Though the BSG technique is presently more mature, the RSG technique has the greatest potential in the era of 30m-class telescopes which will be optimised for near-IR wavelengths and will be equipped with adaptive optics supported multi-object spectrographs. 

RSGs have typical luminosities of  $L \geq$10$^{4}L_{\sun}$  \citep{Humphreys79} and their fluxes peak at $\simeq$1$\mu$m, thus they are extremely bright in the near-IR. In \citet{rsg_jband} we introduced a technique to measure metallicities from $\alpha$-element and Fe lines falling in the 1.15-1.22 $\mu$m window included in the $J$-band. This $J$-band technique has been demonstrated to have several advantages on other methods routinely used to measure metallicity. RSGs are the brightest point sources at wavelengths around 1$\mu$m, and the absence of any significant diffuse emission at these wavelengths means that observations do not suffer from blending and/or \temptext{gas} contamination. Moreover, models predict that the diagnostic lines are easily detectable over a broad range of metallicities from Solar to one tenth Solar \citep{Evans11}, in contrast to auroral (T$_{\rm{e}}$-sensitive, or `direct')  \ion{H}{2} region studies which hit problems at Z $>$ 0.5 Z$_{\sun}$ when the lines become very weak  \citep{Stasinska05,Bresolin05,Ercolano10,Zurita12}.

In \citet{perob1} and \citet{MCpaper} we presented preparatory studies on objects in the Milky Way and Magellanic Clouds, respectively, to validate this new technique. In subsequent works we extended the $J$-band method to extra-galactic distances \citep[][]{Patrick15,Gazak_NGC300,Patrick17}, showing an excellent agreement with other high-precision abundance tracers such as BSGs and auroral  \ion{H}{2} region lines \citep[][]{Gazak_NGC300}. For a summary of the RSG-technique results thus far, see \citet{Messenger}.

Interestingly, we have shown that the $J$-band technique can also be applied to analysis of the integrated light of coeval agglomerates of young stars, i.e. star clusters. In high star-formation-rate galaxies, one can find many such coeval populations with masses in excess of $10^4$\msun, objects known as super-star-clusters (SSCs). Roughly 8 Myr after the formation of the SSC, massive stars which have not yet exploded as supernovae will be in the RSG stage. For a cluster mass of $\sim 10^5$\msun, one may expect $\sim100$ RSGs, and these stars will dominate the near-IR light from the cluster, providing 95\% of the J-band flux \citep[][]{Gazak13}.  As RSGs span only a narrow range of T$_{\rm eff}$, the integrated spectrum can be analysed as the spectrum of an individual RSG star, and the boost in integrated $J$-band flux can be used for quantitative spectroscopy at far greater distances. The potential of this technique was demonstrated by comparing studies of the spectra of individual stars in resolved clusters with that of the clusters' integrated light \citep{perob1,Patrick16}, and exploited for the first time in \citet{Gazak_SSCs} and \citet{Lardo15}. 

In this paper, we have chosen to target M83, the nearest massive, grand-design face-on galaxy, with a Hubble-type SAB(s)c and a distance of 4.9 $\pm$ 0.2 Mpc \citep{Bresolin16}. The galaxy is thought to have high, super-Solar metallicities in its centre, and so is a powerful test-point at the high-metallicity end of the MZR. M83 is known to host a rich SSC population with a wide range of masses and ages \citep[e.g.][]{Bastian12}, one of which we have observed previously,  \citet{Gazak_SSCs}, finding a twice-Solar metallicity ([Z]= 0.28 $\pm$ 0.14 with respect to a Solar $Z_{\odot}=0.012$). Direct abundances, based on the the detection of T$_{\rm e}$ sensitive auroral lines are available for a number of \ion{H}{2} regions \citep{Bresolin05,Bresolin09_M83}, and are in very good agreement with the SSC metallicity measured by \citet{Gazak_SSCs}. More recently, \citet{Bresolin16},  presented a comparative analysis of the metallicities derived from spectra for 14 A-type supergiants in M83 and nebular oxygen abundances from HII regions. They confirmed the super-solar metallcity of M83 and found a good agreement between the stellar metallicities and those inferred from direct T$_{e}$-based methods once a modest correction for dust depletion was applied to the latter. Conversely, many empirically calibrated strong line diagnostics are shown to underestimate the stellar metallicities by $\ga$0.2dex. The O3N2 calibration method by \citet{Pettini-Pagel04} showed the best agreement with the results obtained from blue supergiant stars in the centre of M83.

Here we present metallicity determinations for seven SSCs in M83 from integrated medium resolution NIR spectra, to provide an independent validation of the super-Solar metallicities found from the BSG and nebular studies. We then combine the results of this study with our other recent work in nearby galaxies to construct the first RSG-based MZR.

The plan for the rest of this paper is as follows. Section \ref{sec:obs} presents the observations and target selection. Section \ref{sec:anal} describes the procedures used to derive atmospheric parameters and measure metallicities. We describe our results and compare to other recent work in Sect.\ \ref{sec:results}. In Sect.\ \ref{sec:disc} we assemble our recent results on nearby galaxies, compare to those of BSGs, and compile a mass-metallicity relation for nearby galaxies based entirely on RSG analyses. We conclude in Sect.\ \ref{sec:conc}.

\begin{table*}
  \caption{Coordinates and observed properties of the targets in this study.}
  \begin{center}
    \setlength{\extrarowheight}{3pt}
  \begin{tabular}{ccccccc}
\hline \hline
ID$^{\rm a}$  & RA DEC & $F160W^{\rm a}$ & $V-H^{\rm a}$ & $\log({\rm age/yr})^{\rm a}$ & $\log($M$/\msun)^{\rm a}$ & $R/R_{25}^{\rm b}$ \\
  & (J2000) &   \\ 
\hline
10594 &  13 37 03.4  -29 51 02 & 17.84$\pm$0.03 &  0.25$\pm$0.06 & 6.78 & 4.08  & 0.07 \\
30651 &  13 37 06.2  -29 53 18 & 17.15$\pm$0.04 &  1.32$\pm$0.06 & 7.85 & 5.29  & 0.32 \\
40610 &  13 36 53.7  -29 49 14 & 17.22$\pm$0.02 &  1.10$\pm$0.05 & 7.11 & 4.42  & 0.55 \\
40820 &  13 36 58.1  -29 48 26 & 17.39$\pm$0.02 &  1.52$\pm$0.05 & 6.95 & 4.20  & 0.59 \\
50660 &  13 36 56.9  -29 50 49 & 17.80$\pm$0.01 &  1.64$\pm$0.05 & 7.30 & 4.28  & 0.26 \\
60571 &  13 36 52.5  -29 53 16 & 16.94$\pm$0.03 &  0.65$\pm$0.06 & 6.80 & 4.28  & 0.38 \\
60596 &  13 36 55.1  -29 53 14 & 17.44$\pm$0.03 &  1.36$\pm$0.06 & 7.48 & 4.72  & 0.30 \\
\hline
  \end{tabular} \\
  \end{center}
  $^{\rm a}$ Data from \citet{Bastian11}. \\
 $^{\rm b}$ Assuming the centre of M83 is at $\alpha=13h 37m 0.9s, \delta=-29\degr 51^\prime 57^{\prime\prime}$, and $R_{25} = 6.44^\prime$ \citep{deVaucouleurs91}.
  \label{tab:obs}
\end{table*}

\begin{figure*}
\begin{center}
\includegraphics[width=17cm]{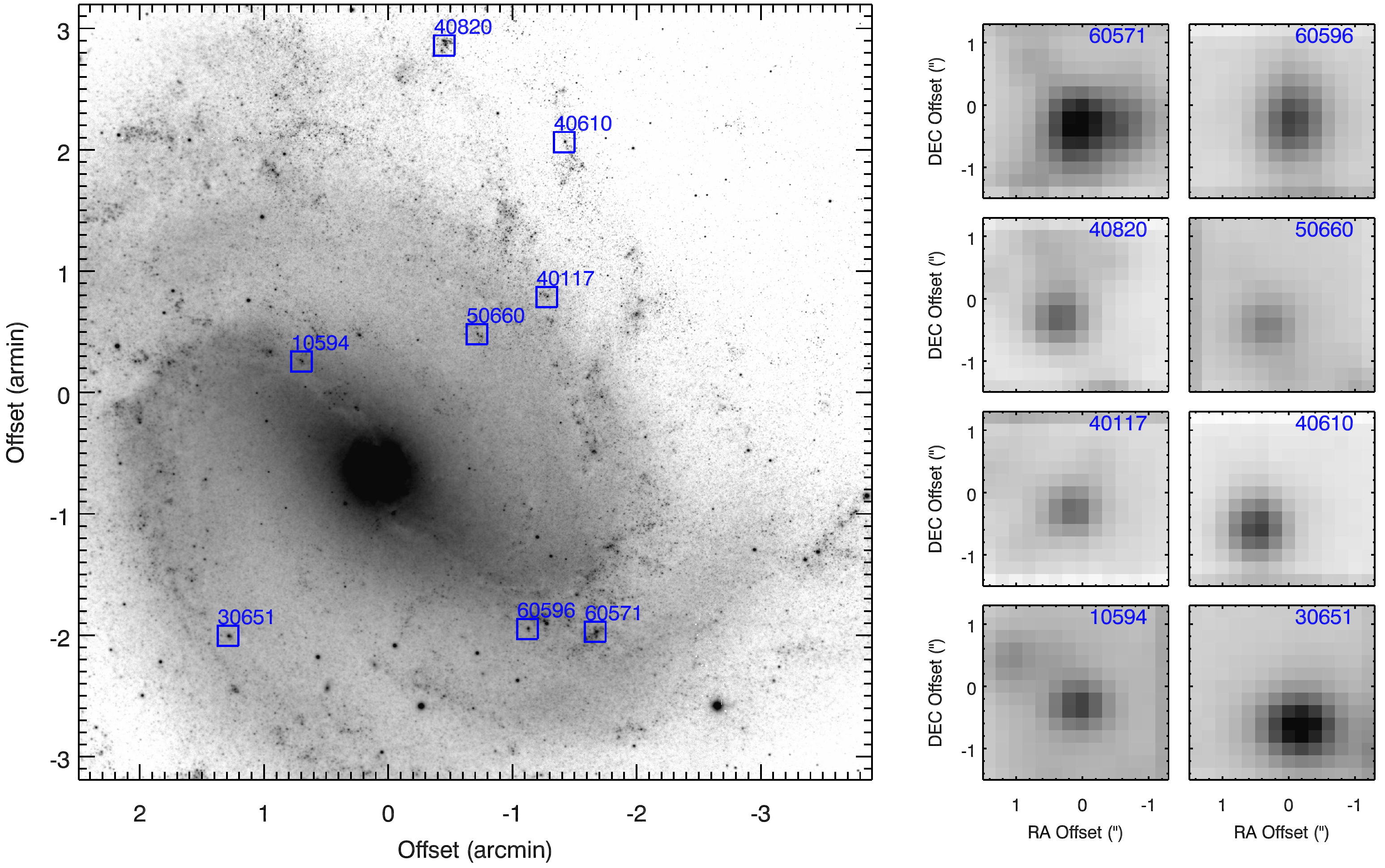}
\caption{{\it Left}: $J$-band image of M83, centred on the KMOS field centre ($\alpha=$13:36:56.97; $\delta=$-29:51:17.9 ), with the observed clusters indicated by blue boxes. {\it Right}: images of each of the clusters, made from median collapsing the KMOS IFU datacubes. }
\label{default}
\end{center}
\end{figure*}

\section{Observations \& data reduction} \label{sec:obs}

For our target selection, we begin with the list of SSCs in \citet{Bastian11,Bastian12}, studied also in \citet{Gazak_SSCs}. We select a subsample of objects with derived ages below 100Myr but with F110W-F160W colours greater than zero (to ensure the presence of RSGs), with F160W magnitudes $<$18 to target the most massive clusters with lots of RSGs. The final sample of targets, along with their observed parameters, is listed in Table \ref{tab:obs}. 

The data for this paper was taken with KMOS, the near-infrared multi-object spectrograph at the VLT. The data was taken during the nights of the 16th and 18th April 2016 in partly cloudy conditions as part of the observing programme 097.B-0281 (PI: C.\ Evans).  The KMOS instrument has 24 deployable arms each which feed one integral field unit (IFU). The data from the IFUs is then fed to three detectors, with eight IFUs stored on each. Our observing strategy was to assign two IFUs to each target, one on the object and one on a nearby patch of blank sky, and nod between them in an ABAB pattern. This ensures that there is no downtime for sky observations, at the expense of reducing the maximum possible multiplexing from 24 to 12. We specified detector integration times (DITs) of 150sec and NDIT=2 before nodding between IFUs. This time between nods of 300sec was selected in order to be short enough such that the airglow emission lines do not change substantially between nods, whilst being long enough to obtain good signal-to-noise (S/N) on the airglow lines, essential for our data-reduction process (see below). In total, we had 57 repeated observations of each target, giving a total on-target integration time of 4.75hrs. 

In addition to the science target, we observed the star HIP66419, spectral type A0V, as a telluric standard. This star was observed at least every 2 hrs, to ensure that any science target observation was no more than 1hr apart from a telluric observation. The standard star was observed through every IFU allocated to a science observation, using the {\tt std\_star\_scipatt} template. The standard suite of calibration observations (darks, flats, sky flats, wavelength calibrations) were taken during the daytime. 

The initial stages of the data-reduction process were performed using the KMOS pipeline \citep{KMOSpipeline}. This included the preparation of the darks and flats, wavelength calibration, illumination correction and reconstruction of the datacubes for each observation though each IFU. This reconstruction was done specifying the sampling to be 3 pixels per resolution element (i.e. by setting the keyword {\tt b\_samples=3072}). This is greater than the instrumental sampling, which is 2 pixels per resolution element. At the instrument's level of sampling, applying sub-pixel shifts to e.g. telluric standard spectra can easily introduce extra numerical noise. Since resampling the spectra during the reconstruction phase is inevitable and unavoidable due to the non-linear wavelength correction, we have found that overall numerical noise is best minimised by slightly super-sampling the spectra at the reconstruction stage. The individual science spectra are then never resampled again. 

The methodology of the second phase of the reduction is described in \citet{Gazak_NGC300,Lardo15} and \citet{Messenger}. It is known that the wavelength calibration of KMOS varies across each IFU, between IFUs, between detectors, and can change in time throughout the night as the instrument rotates. To correct for these effects, the sky emission lines were used to fine-tune the wavelength calibration at each spaxel, as well as determine the spectral resolution at a specified central wavelength (chosen to be the centre of our analysis window, 1.18\um). Spectra of the science targets were extracted using a narrow aperture of radius 1.5\arcsec\ to minimise the effect of the spatially varying spectral resolution. It is important to note that the spectra are {\it not} resampled on to the updated wavelength axis, since this can introduce numerical noise. Instead, each spaxel's new wavelength axis is stored as a separate array until the combination stage (see below). 

From each science spectrum we subtracted the corresponding sky spectrum, taken with the same IFU in the nod position. We adjusted the sky emission lines by scaling them by a factor optimized to provide the best cancellation (a process we call `sky-tuning'), concentrating on the spectral window used in our analysis ($\sim$1.15-1.22\um). The optimum scaling factor was typically very small, $< \pm 5\%$, since the time between target and sky observations was just 5 mins. 

Each science spectrum was then corrected for telluric absorption by dividing through by the telluric standard star. The cancellation of the telluric features was again optimised automatically, this time by tuning the relative wavelength shift, the spectral resolution and the strength of the telluric lines relative to the continuum. The optimum values of each of these quantities were determined by searching for the combination that yields the lowest variance in the corrected science spectrum across our spectral window. 

The repeated observations for each science target were then combined. Since each individual observation has a slightly different wavelength axis, we first defined a master wavelength axis with a similar pixel scale, which we call $\Delta \lambda$. The value of $\Delta \lambda$ is defined to be one third of a resolution element of the lowest resolution spectrum in the individual observations of the same target. For each spectral pixel $\lambda_i$ on the master axis, we median-combined the fluxes of all the individual observations between $\lambda_i \pm \Delta \lambda / 2$. The error on the flux at $\lambda_i$ is taken to be $\sigma_i / \sqrt(n)$, where $\sigma_i$ is the standard deviation of the fluxes at $\lambda_i$ and $n$ is the number of individual observations, in this case 57. We found that this empirical estimate of the noise is the most reliable, since it takes into account not only photon shot-noise but also variations in the quality of the correction to the airglow and telluric lines. The S/N for each of our targets was greater than 80 per pixel, with the exception of \#50660 which has a S/N $\sim$50-60 in our spectral window of 1.15-1.22\um.

\begin{figure*}
\begin{center}
\includegraphics[width=17cm]{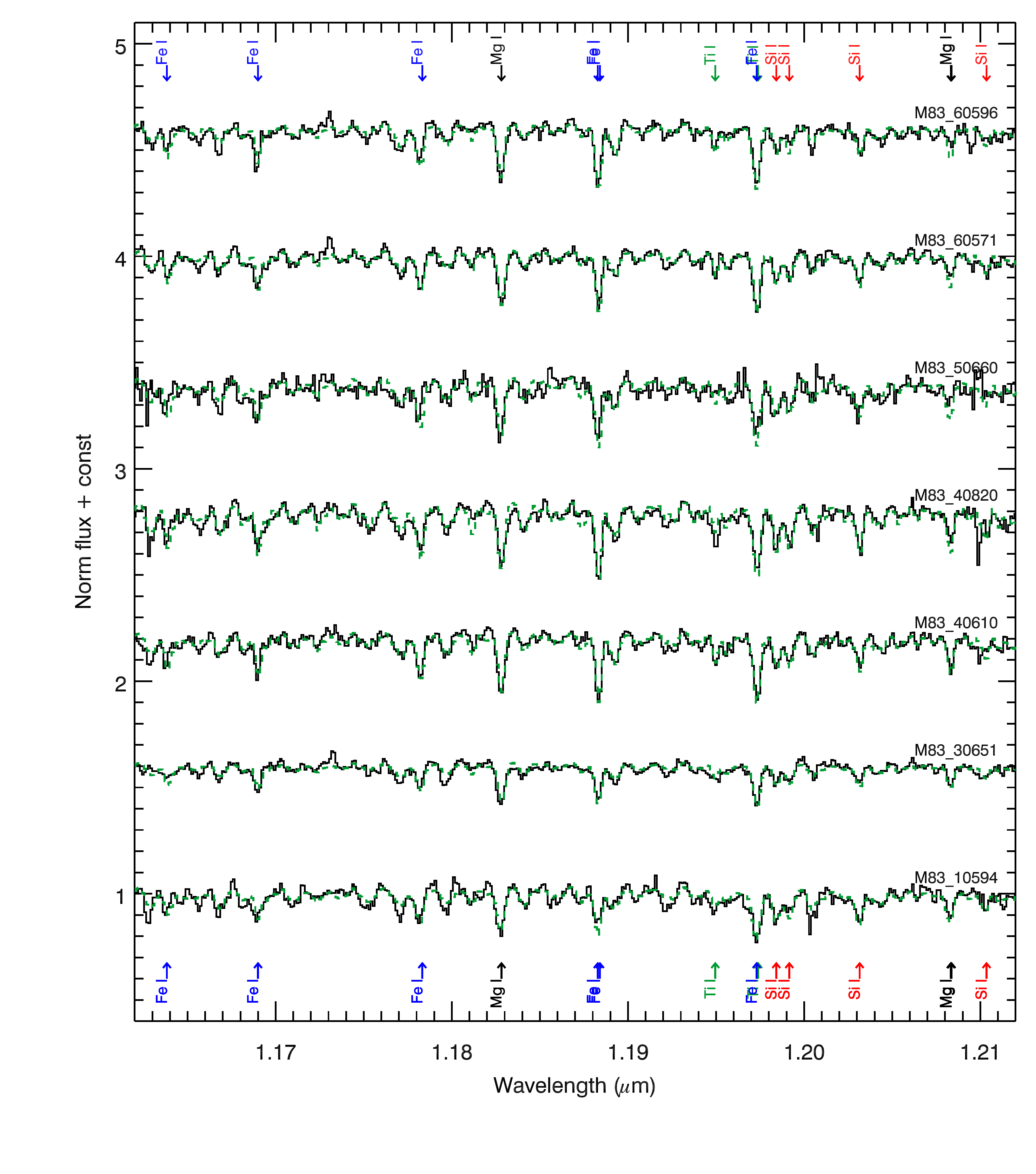}
\caption{$J$-band spectra of the clusters in this study (black solid lines), and their best-fitting models (green dashed lines). Prominent spectral lines have been indicated.}
\label{fig:bestfits}
\end{center}
\end{figure*}

\section{Analysis} \label{sec:anal}

Our analysis methodology follows that first described in \citet{rsg_jband}, and later in more detail in \citet{MCpaper}. As described in Sect.\ \ref{sec:intro}, once the age of an SSC passes $\sim$8Myr its near-IR light is dominated by the RSGs. Since RSGs with the same metallicity appear almost identical in the spectral window 1.15-1.22\um, the spectrum of the SSC is indistinguishable from that of a single star. We can therefore analyse the spectrum of each SSC as if it were a single RSG. This approach has been shown to work well from analysis of nearby star-clusters where we can resolve the individual stars \citep{perob1,Patrick16}. 

To analyse the clusters' spectra, we have computed a grid of MARCS model atmospheres \citep{gustafsson08} that span the parameter ranges appropriate for RSGs: effective temperatures 3400K$\le$\teff$\le$4400K (steps of 100K); gravities -1.0$\le$\logg$\le$+1.0 (steps of 0.5dex); metallicities relative to Solar -1.0$\le$[Z]$\le$+1.0 (steps of 0.25dex); and microturbulent velocities 1$\le \xi$ \kms$\le$6 (steps of 1\kms). The Solar abundances are taken from \citet{asplund05}, corresponding to a Solar metal fraction of $Z_{\odot} = 0.012$. We assume a Solar ratio of $\alpha$-elements to Fe\footnote{\temptext{Though we have both Fe and $\alpha$-element lines in our spectra, we do not attempt to measure the $\alpha$/Fe ratio. Adding this fifth free parameter substantially increases the degeneracy errors since the ratio of the strengths of the Fe and Si lines are also sensitive to \teff, owing to the large difference in the excitation potentials of the lines of these elements. Incorporating other Fe lines into our analysis with different excitation potentials, which would permit a measurement of $\alpha$/Fe, is work in progress.}}. From these models, we compute synthetic spectra using an updated version of the SIU code \citep{Bergemann12}. The spectra are computed in local thermodynamic equilibrium, but crucially the dominant species in the $J$-band spectral window (\fei, \mgi, \tii, \sii) are computed in non-LTE as described in \citet{Bergemann12,Bergemann13,Bergemann15}. The spectra are computed at very high resolution ($R=500,000$), and are then degraded to the spectral resolution of the observations, which was determined at the advanced stages of the data-reduction process (see Sect.\ \ref{sec:obs}). 

The spectrum to be analysed is first continuum-normalised. We divided through by a template model spectrum which has had the diagnostic lines masked out, then smoothed heavily with a median filter, then fit a high-order polynomial to the resulting smoothed ratio spectrum. We choose a low-metallicity template spectrum so that its continuum is clearly definable, though the exact choice of this template spectrum makes no difference to the results since we then normalize the grid of model spectra in the same way. The strengths of the diagnostic lines in the data are measured by fitting gaussian profiles using the IDL function {\tt gaussfit}. By including the error spectrum, we also obtain the uncertainties on these line-strengths. The strengths of the lines in the model grid are then measured in exactly the same way, including the same continuum fitting process. 

The best-fitting model is found by a \chisq-minimisation process, matching the strengths of the diagnostic lines in the data to those in the model grid. The values of the best-fitting parameters are determined from the location in the model grid where the \chisq\ derivative goes to zero, and the uncertainties on each parameter are estimated by considering all models that have \chisq values within $\Delta \chi^2$=2.3 of the minimum. We also add in quadrature those errors propagated from the uncertainty on the spectral resolution $R$, though in practice these are negligible since $R$ is well-constrained to a precision of $\sim2\%$ from measurements of the sky airglow lines. Finally, in order to compare our metallicity measurements to those from BSGs and \hii-regions, we need to correct for a small effect arising from the fact that our RSG models use the \citet{asplund05} element abundance pattern, whereas the low resolution analysis of extragalactic BSG stars assumes the Asplund et al. (2009) value of $\log(O/H)+12=8.69$ for the solar oxygen abundance and the values from Grevesse and Sauval (1998) for all other elements. We therefore subtract 0.084dex from our logarithmic metallicities in order to be on the same scale with the metallicities obtained from the BSG analysis (see Appendix \ref{sec:app} for details).

\begin{figure}
\begin{center}
\includegraphics[width=8.5cm]{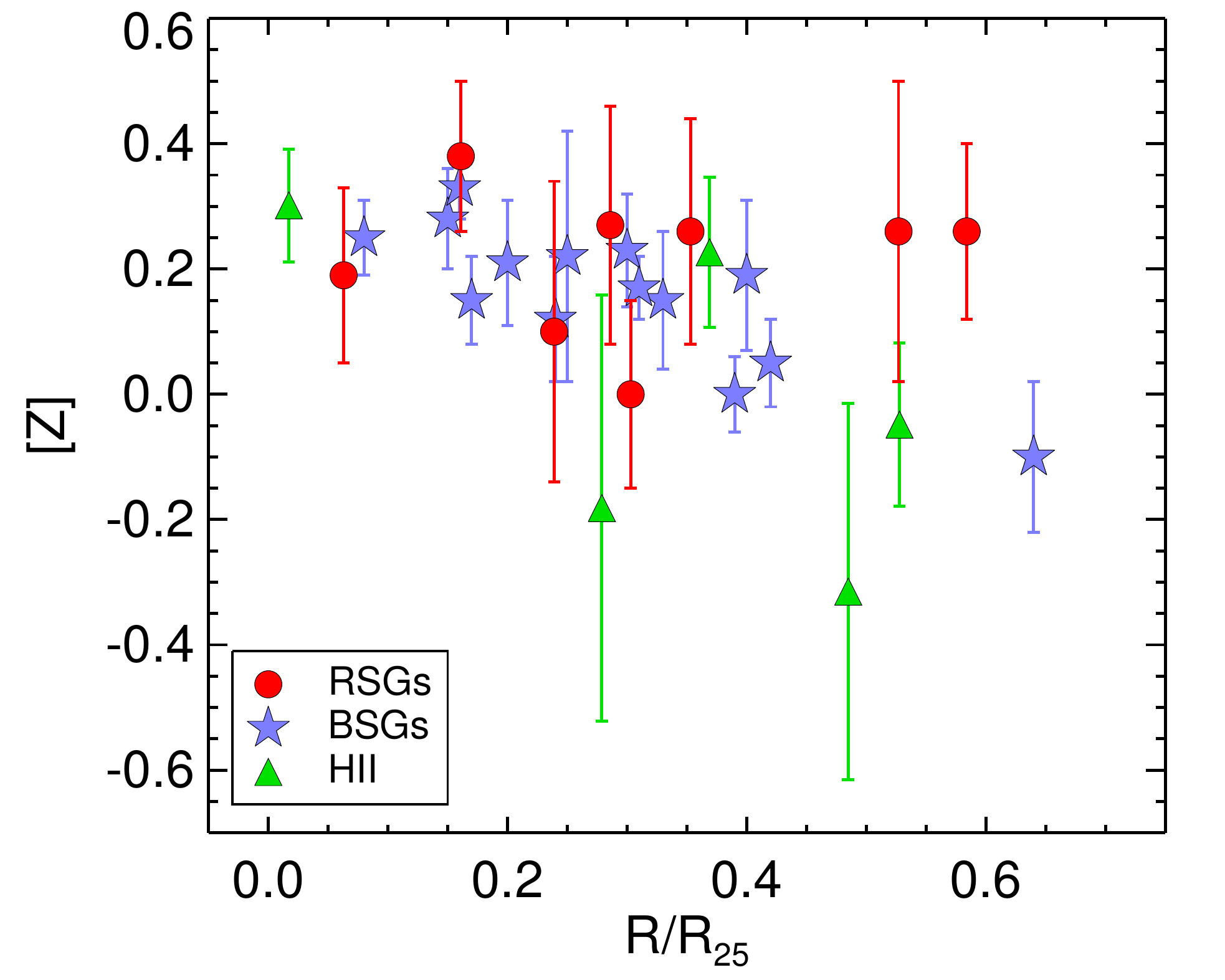}
\caption{{The metallicity M83 as a function of galactocentric distance measured in units of the isophotal radius. The red circles are data from this work, apart from the point at the lowest $R/R_{25}$ which is the SSC from \citet{Gazak_SSCs}.  The blue stars are measurements of blue supergiants by \citet{Bresolin16}. Also included are the direct-method HII-region metallicities (green triangles).}}
\label{fig:zgrad}
\end{center}
\end{figure}


\begin{table}
  \caption{Best-fitting model parameters for each of the targets in this study. }
  \setlength{\extrarowheight}{6pt}
  \begin{tabular}{lcccc}
\hline \hline
Star & \teff\ (K) & log(g) & $\xi$ (km/s) & [Z] \\
\hline
10594 & 4040 $\pm$  80 &  0.3 $\pm$ 0.3 & 1.5 $\pm$ 0.3 &  0.38 $\pm$ 0.12 \\
30651 & 3850 $\pm$  80 &  0.5 $\pm$ 0.2 & 1.2 $\pm$ 0.3 &  0.00 $\pm$ 0.15 \\
40610 & 3780 $\pm$ 110 &  0.4 $\pm$ 0.2 & 2.7 $\pm$ 0.3 &  0.26 $\pm$ 0.24 \\
40820 & 4170 $\pm$ 130 &  0.5 $\pm$ 0.2 & 3.0 $\pm$ 0.4 &  0.26 $\pm$ 0.14 \\
50660 & 4080 $\pm$ 120 &  0.4 $\pm$ 0.2 & 3.5 $\pm$ 0.5 &  0.10 $\pm$ 0.24 \\
60571 & 3820 $\pm$  70 &  0.4 $\pm$ 0.2 & 1.7 $\pm$ 0.2 &  0.26 $\pm$ 0.18 \\
60596 & 3630 $\pm$  80 &  0.5 $\pm$ 0.2 & 2.9 $\pm$ 0.4 &  0.27 $\pm$ 0.19 \\
\hline
  \end{tabular}
  \label{tab:results}
\end{table}

\section{Results} \label{sec:results}

The spectra of each of the target clusters, along with their best-fitting spectra, are displayed in \fig{fig:bestfits}. The best-fit model parameters are listed in Table \ref{tab:results}. We find a weighted average metallicity for all clusters of $[Z]=0.21\pm0.11$ relative to a Solar value of $\zsun=0.014$. There are no systematic trends with SNR or spectral resolution. \temptext{We repeated the abundance analysis with different versions of the reduced spectra, for example those which have had the the sky-tuning or kmogenization algorithms activated or de-activated. Each version of the reduction yielded spectra with average metallicities consistent to within $\pm0.05$dex, though with larger dispersions for those reductions without sky-tuning or kmogenization}.

In two of the clusters observed (10594, 60571), as well as the usual RSG spectral features we also detected \hei\ 10\,830\AA\ in emission. In one more cluster (60596) we saw what looked like a weak \hei\ 10\,830\AA\ absorption line. This may indicate the presence of hot massive stars, and hence is suggestive of young ages ($\la 10$Myr), consistent with the bluer colours for first two of these clusters (see Table \ref{tab:obs}). As shown by \citet{Gazak_SSCs}, at very young ages the contribution to the total $J$-band flux by the hot stars may be non-negligible, causing the RSG spectral features to be diluted and potentially resulting in an underestimate of the metallicity. However, the clusters with \hei\ tend to be those with the highest metallicity. We therefore see no obvious evidence of systematic offsets in metallicity for these clusters. On contrast, one of the clusters we observed (40117) had strong \hei\ 10\,830\AA\ emission but a complete lack of RSG features, suggesting a very young age for this object. We therefore excluded this cluster from our analysis. 

The radial metallicity distribution of M83 determined from our observations of the SSCs is displayed in \fig{fig:zgrad}. We also include in the plot the SSC close to the centre of M83 studied in \citet{Gazak_SSCs}. From the SSCs alone there is no obvious metallicity gradient within $R/R_{25} = 0.6$, the formal gradient measured from the SSCs is $-0.04\pm0.32$dex per $R_{\rm 25}$, and so is consistent with zero. Again, this measurement is robust to the details of the data reduction.  

In \fig{fig:zgrad} we also compare the SSC metallicities with those measured from blue supergiants (BSGs), published recently in \citet{Bresolin16}. Inside $R=0.5R_{25}$ the two samples look similar, with averages that are consistent, $[Z_{\rm BSG}] = 0.18\pm0.09; [Z_{\rm SSC}] = 0.20 \pm 13$, with the two sets of points not showing dissimilar trends. Once the outlying points with $R > 0.5R_{25}$ are considered, there seems to be a greater discrepancy between the two samples, though the differences are within 2$\sigma$ and are caused by the outermost BSG, which seems to be an outlier. \temptext{We note that the differences between the BSGs and the SSCs at larger radii {\it cannot} be due to azimuthal variations, since the outermost BSG in \fig{fig:zgrad} \citep[$\equiv$ \#5 in][]{Bresolin16} is only $\sim$20\arcsec\ from SSC \#40820.}

Also overplotted in \fig{fig:zgrad} are the direct-method \hii-region datapoints from \citet{Bresolin05}, which were updated in \citet{Bresolin16}. Though there are only five points, and the errors are large, the agreement with both RSG and BSG measurements is satisfactory. Though the outer BSG and \hii-region points hint at a metallicity gradient, the BSG-calibrated strong-line \hii-region measurements indicate that there is no discernible gradient within $R=0.8R_{25}$ \citep{Bresolin16}.



\begin{table}
\caption{The nearby galaxies with supergiant metallicity measurements at a galactocentric radius of 0.4$R_{25}$. }
\begin{center}
  \setlength{\extrarowheight}{6pt}
\begin{tabular}{lcccl}
\hline \hline
Galaxy & $\log(M_{\star}/M_{\odot})$ & [Z]$_{\rm RSG}$ & [Z]$_{\rm BSG}$ & Ref.\ \\
\hline
     M31 & 11.0 &  --   &  0.07 & 1\\
     M81 & 10.9 &  --   &  0.12 & 2\\
      MW & 10.8 &  0.04 &  0.11 & 3,4,5 \\
     M83 & 10.6 &  0.21 &  0.16 & This work, 6\\
 NGC4083 & 10.5 & -0.01 &  --  & 7\\
 NGC3621 & 10.3 &  --   & -0.01 & 8 \\
     M33 &  9.6 &  --   & -0.20 & 9 \\
   NGC55 &  9.3 & -0.52 & -0.46 & 10,11 \\
     LMC &  9.2 & -0.45 & -0.36 & 12,13 \\
  NGC300 &  9.0 & -0.29 & -0.24 & 14,15 \\
     SMC &  8.7 & -0.61 & -0.65 & 12,13 \\
 NGC6822 &  8.2 & -0.60 & -0.52 & 16,17 \\
 NGC3109 &  8.1 &  --   & -0.67 & 18 \\
  IC1613 &  8.0 &  --   & -0.79 & 19 \\
     WLM &  7.7 &  --   & -0.87 & 20 \\
    SexA &  7.4 &  --   & -1.00 & 21 \\
\hline
\end{tabular}
\end{center}
\label{tab:mzr}
{\bf Refences}: 
1: \citet{Przybilla08}; 
2: \citet{Kudritzki12}; 
3:\citet{rsg_jband}; 
4:\citet{perob1}; 
5: \citet{Przybilla06};
6:\citet{Bresolin16}; 
7: \citet{Lardo15}; 
8: \citet{Kudritzki14};
9: \citet{U09};
10: \citet{Patrick17}; 
11: \citet{Kudritzki16}; 
12: \citet{MCpaper}; 
13: \citet{Hunter07};
14: \citet{Gazak_NGC300}; 
15: \citet{Kudritzki08};
16: \citet{Venn01}
17: \citet{Patrick15}; 
18: \citet{Hosek14};
19: \citet{Bresolin07};
20: \citet{Urbaneja08};
21: \citet{Kaufer04};
\end{table}%

\begin{figure}
\begin{center}
\includegraphics[width=8.5cm]{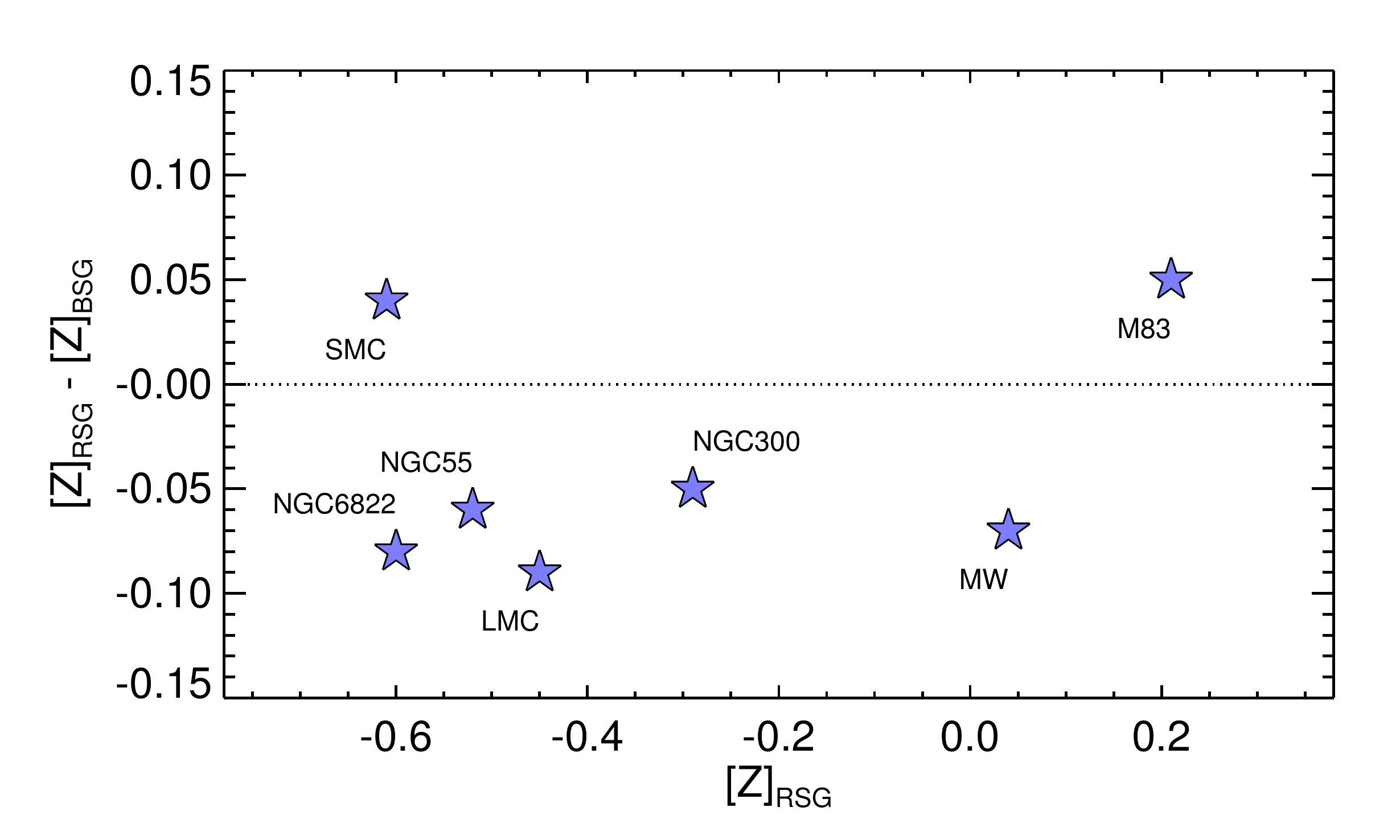}
\caption{A comparison of the average integrated metallicities derived for all galaxies with both a RSG and a BSG measurement. See Table \ref{tab:mzr} and Sect.\ \ref{sec:rsg-bsg} for details.}
\label{fig:rsgs-bsgs}
\end{center}
\end{figure}

\begin{figure*}
\begin{center}
\includegraphics[width=8.5cm]{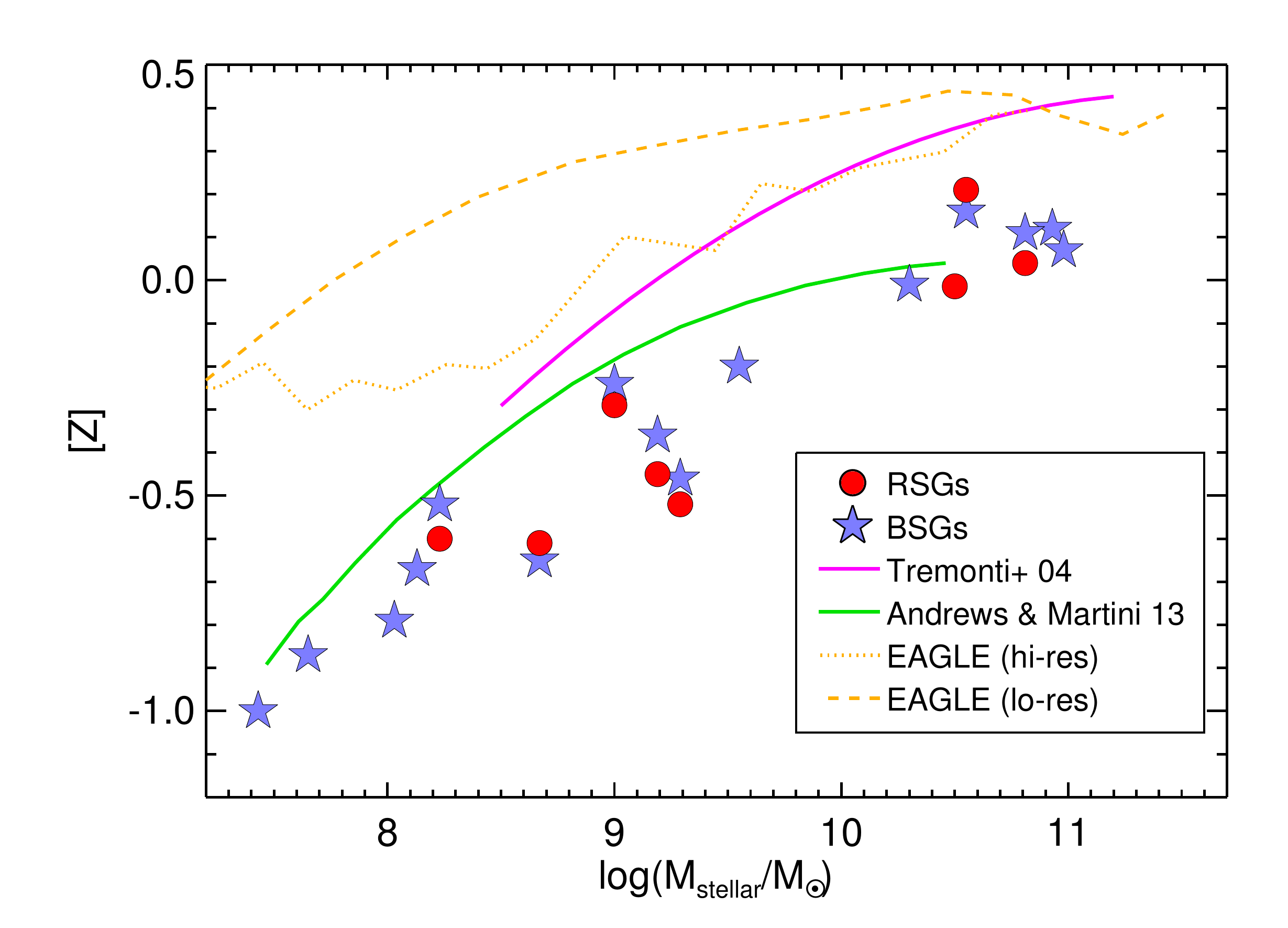}
\includegraphics[width=8.5cm]{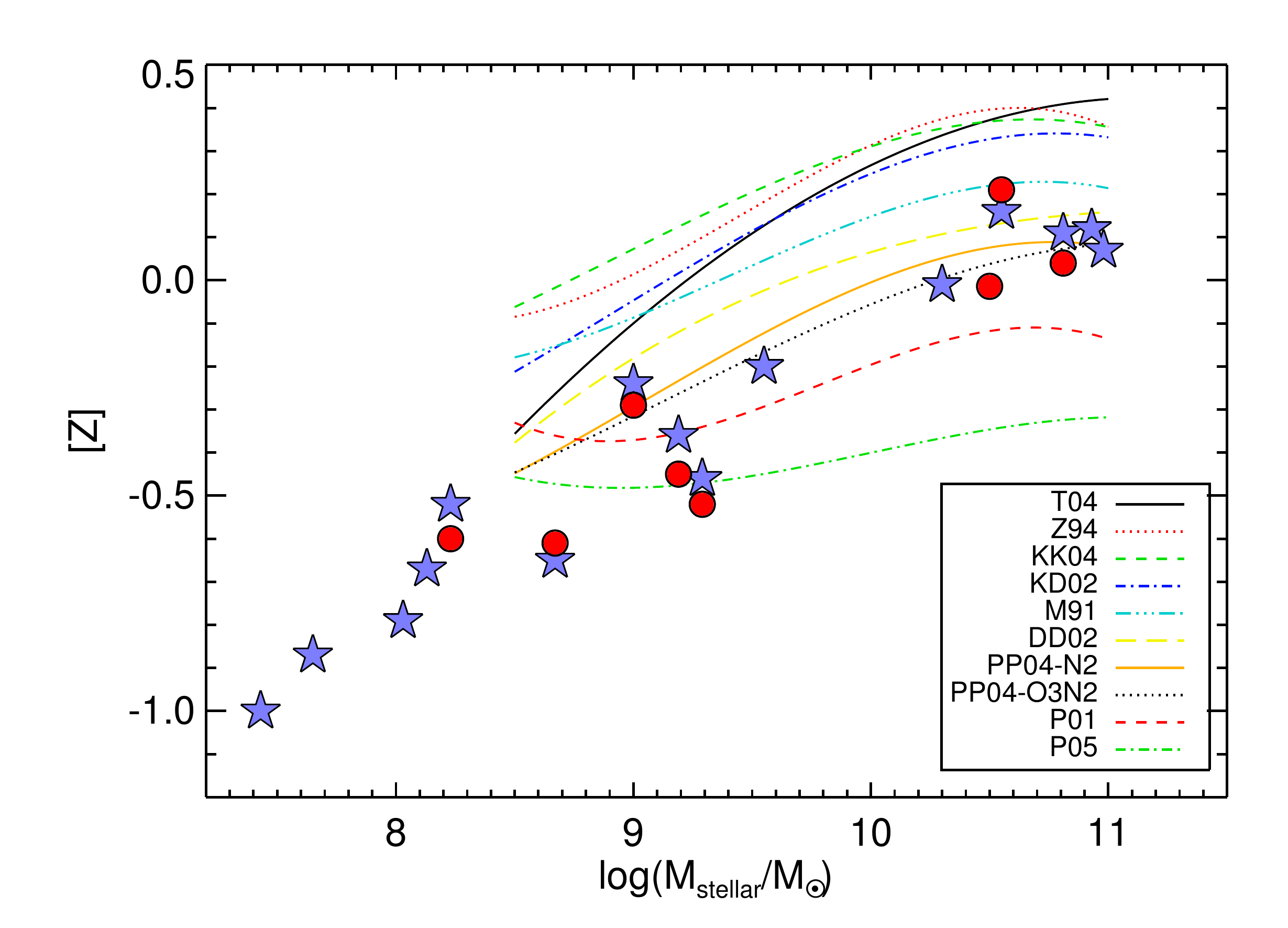}
\caption{The mass-metallicity relation as measured from red/blue supergiants (red circles and blue stars respectively). In the left plot, we compare to the strong-line calibrated \hii-region study of SDSS galaxies by \citet{Tremonti04}, the binned direct-method study of the same sample by \citet{Andrews-Martini13}, and the results of the EAGLE simulation. In the right plot, we compare the supergiant MZR to various strong-line calibrations of the SDSS data in Tremonti et al. }
\label{fig:mzr}
\end{center}
\end{figure*}

\section{The Mass-Metallicity Relation of Nearby Galaxies}  \label{sec:disc}

Including this present study on M83, we have now obtained RSG-based metallicity estimates on eight galaxies spanning 3 orders of magnitude in stellar mass. Since the definition of a galaxy's characteristic metallicity is somewhat subjective, we now discuss how we define this quantity. For three galaxies in our sample (LMC, SMC, NGC4083) there is no significant metallicity gradient, and the characteristic metallicity is simply the average of all individual targets (whether stars or clusters) within that galaxy. For those galaxies that do have detectable abundance gradients (M81, MW, NGC300, M33, NGC55) we have specified that the `central metallicity' is that at a galactocentric distance of $0.4R_{25}$, so as to be comparable to the integrated metallicity determined from an SDSS fibre \citep[following][]{Moustakas-Kennicutt06}\footnote{For the MW, we take our measurement of the metallicity in roughly the Solar neighbourhood (at a galactocentric distance of 8kpc), then extrapolate to $0.4R_{25}=5.3$kpc \citep{Goodwin98} assuming a metallicity gradient of -0.06dex/kpc \citep{Genovali14}. }. The results for each galaxy are presented in Table \ref{tab:mzr}. The masses of each galaxy are taken from \citet{Bresolin16} and references therein, with the exception of NGC~4083 where we adopt a stellar mass of $10^{10.5}$\msun\footnote{The mass for NGC~4083 is determined from the total $B$-band luminosity of the Antennae of $2.9\times10^{10}$\lsun\ \citep{deVaucouleurs91}, a $B$-band mass-to-light ratio of between 0.5-3.0 \citep{Bell-deJong01}, and assuming that NGC~4083 contributes roughly half the mass of the system. The experimental uncertainty on this galaxy's mass is around $\pm$0.2-0.3dex.}. Note that this galaxy's mass is approximately half of the total mass of the Antennae system. 

Of the eight galaxies in our sample, two have metallicities derived from star clusters as opposed to individual stars (M83 studied in this paper, NGC4083 in Lardo et al.\ 2015), which may be a potential source of systematic error. However, we note that for two other galaxies (MW and LMC) we have measurements of both field stars {\it and} star clusters, and that we resolve the individual stars in these clusters. We have shown in previous work that (a) the metallicities of the individual cluster members agree with that from the cluster's integrated light; and (b) that the metallicities of the star clusters matches that of the field stars in the same galaxy \citep{rsg_jband,perob1,Patrick16}. Therefore, we expect any systematic offset between cluster and individual star measurements to be negligible with respect to the measurement errors. 

The formal errors on each metallicity measurement in Table \ref{tab:mzr} is small; the standard deviation is typically $\la$0.2dex, with $\sim$10 objects per galaxy, meaning that the error on the mean is $\la$0.07dex. At this level of precision we would expect systematic errors to become important. Before we turn to the MZR, we first assess these systematics by comparing our metallicities with those from BSG-based studies of the same galaxies. 

\subsection{Comparison of RSG and BSG metallicity estimates} \label{sec:rsg-bsg}
Recall that, though the RSG and BSG methods are both based on stellar spectroscopy, they each rely on completely different model atmospheres and diagnostic lines. The BSG method requires hot ($\> 10$kK) model atmospheres, and metal abundances are constrained by singly- and doubly- ionized lines in the optical. In contrast, the RSG technique uses cool ($<$4500K) model atmospheres and employs neutral metallic lines in the near-IR to get metallicities. We therefore consider the RSG and BSG methods to be completely independent of one another. 

There are seven objects in common between the RSG and BSG work, their results are listed in Table \ref{tab:mzr}. In \fig{fig:rsgs-bsgs} we compare the two metallicity estimates for these galaxies. The plot shows that the agreement between the two methods is excellent. Formally, the mean offset between the two $\Delta [Z] \equiv [Z]_{\rm RSG} - [Z]_{\rm BSG}$ = -0.037$\pm$0.058. The systematic offset between the two is therefore consistent with zero, with a standard deviation close to that expected from random experimental errors alone (see previous section). 

The relative agreement between the RSG and BSG methods presented here is similar to that found by \citet{Gazak_NGC300}, who studied the internal metallicity variations in NGC~300. In this galaxy it is also possible to compare the supergiant metallicities to a third independent metallicity diagnostic, that of direct-method \hii-region analyses \citep{Bresolin09}. Gazak et al.\ showed that each method produces a nearly identical abundance gradient [$-0.083 \pm 0.014; -0.081 \pm 0.011; -0.077 \pm 0.006$ dex\,kpc$^{-1}$] , with average metallicities at $R_{25}$ = [-0.50; -0.47; -0.53] for  RSGs, BSGs and \hii-regions respectively. The extremely high level of consistency between the RSG and BSG methods for galaxies spanning metallicities from [Z]=-0.6 to +0.2, plus the excellent agreement between RSGs, BSGs and \hii-regions within NGC~300 spanning [Z]=-0.6 to 0.0, gives us great confidence in each method's absolute precision.

\subsection{Mass-Metallicity Relation}
In \fig{fig:mzr} we plot the MZR as determined from all galaxies studied thus far. The plot again serves to show the excellent agreement between the RSG and BSG techniques. In the left panel of the plot we compare to the results from the survey of $\sim$40,000 SDSS galaxies presented in \citet{Tremonti04}, where we see a substantial systematic offset of around +0.4dex with respect to the supergiants at all metallicities. The cause of this offset is almost certainly systematic errors in the strong-line calibration employed by Tremonti et al., which are well known \citep[e.g.][]{K-E08}. 

Overplotted in green in \fig{fig:mzr} is the work of \citet{Andrews-Martini13}, who took the same data used by \citet{Tremonti04} but binned the galaxy spectra together according to stellar mass and star-formation rate. With the resulting improvement in signal-to-noise, the auroral lines could be detected, and so direct-method metallicities could be obtained. These results are overplotted as the green line in the left panel of \fig{fig:mzr}. Here we see that the offset between the \hii-region and supergiant results is much reduced with respect to Tremonti et al. 

Also, in \fig{fig:mzr} we compare to the MZR from the EAGLE simulation \citep{Schaye15}. The results of two simulations are plotted; that from a large volume at lower mass resolution, and that from a smaller volume at higher resolution (for more details on these simulations, the reader is directed to the Schaye et al.\ paper). We compare to the EAGLE `gas phase' metallicities rather than the mass-weighted stellar metallicity. The latter will be heavily skewed towards the older ($\ga$1Gyr) stellar population, and the parent galaxy may well have undergone significant chemical evolution since these stars formed. In contrast, supergiants are very young in cosmological terms ($\sim$10-50Myr), and their host galaxy's interstellar medium will have undergone very little chemical evolution since these stars were formed. Therefore, we expect the average metallicity of a galaxy's blue and red supergiants to be directly comparable to that of its star-forming (i.e.\ \hii) regions. The plot shows that, regardless of the simulation type, there are substantial offsets between the simulation results and the supergiants survey at all galaxy masses, particularly at the low mass end where the discrepancy is $\ga 0.5$dex. 

In the right panel of \fig{fig:mzr} we compare our supergiant-based MZR to those obtained by applying various strong-line calibrations to the same SDSS data presented in \citep{Tremonti04} \citep[following][]{K-E08}. From visual inspection, the best match to the supergiant data is obtained using either of the \citet{Pettini-Pagel04} calibrations (N2 or O3N2 see Pettini \& Pagel for their definitions). This is consistent with that found for the metallicity gradients in NGC~300 and M83 \citep{Bresolin09,Bresolin16}, who find the best agreement with the O3N2 calibration. The BSG and RSG results therefore verify the accuracy of the O3N2 diagnostic at high metallicities, a regime where the original calibration relied upon sparse sampling and photo-ionization modelling rather than purely direct methods.

\section{Summary} \label{sec:conc}
We have presented a metallicity study of the central regions of M83 using a sample of seven Red-Supergiant (RSG) dominated star-clusters. The flat abundance gradient at a level of approximately twice Solar metallicity found in this work agree well with that derived from blue supergiants (BSGs) and from direct-method \hii-region studies. Since these three methods are completely independent, we interpret this result as strong evidence that each method provides metallicities accurate on an absolute scale to within $\sim$0.05dex. 

We have also compiled all of our recent RSG-based metallicity studies to study the relationship between mass and metallicity for a sample of nearby galaxies. Again we find excellent agreement between the RSG and BSG methods for the seven galaxies in common, with a dispersion of $\pm$0.06dex and a systematic offset consistent with zero. The supergiant-based mass-metallicity relation is systematically offset from the Tremonti et al.\ (2004) measurement from SDSS, the former being $\sim$0.4dex lower. We interpret this as being due to the well-known problems with strong-line \hii-region methods. Indeed, we find much better agreement with the `direct-method' \hii-region study of Andrews \& Martini (2013) obtained by binning the SDSS spectra. We find that the strong-line calibration yielding the most accurate metallicities is the O3N2 calibration of Pettini \& Pagel (2004), which appears to hold between metallicities of SMC-like to 2-3$\times$ Solar.

\acknowledgments Acknowledgments: The data used in this paper is from ESO programme 097.B-0281 (PI: C.\ Evans). Part of the work in this paper resulted from the workshop {\it The Chemical Evolution of Galaxies}, hosted by the Munich Institute for Astro- and Particle Physics (MIAPP) in 2016. BD acknowledges financial support from the Science and Technology Research Council (STFC). This work was supported in part by the National Science Foundation under grant AST-1108906 to RPK. For our analysis we used the software package IDL, The IDL Astronomy User's Library at GSFC, and the Coyote graphics library.

\bibliographystyle{/fat/Data/bibtex/astronat/apj/apj.bst}
\bibliography{/Users/astbdavi/Google_Drive/drafts/biblio}

\appendix
\section{A: Homogenizing the RSG and BSG abundances} \label{sec:app}
In our RSG work, we are primarily sensitive to the absolute metal fraction, $Z$. Though there may be a minor dependence on the He/H ratio due its effect on the temperature structure of the MARCS model atmospheres and on the H$^-$ continuum, we expect these effects to be minor. Our work adopts the metal fractions of \citet{asplund05}, which has $Z = 0.012$ and $z^\prime \equiv Z/X = 0.0165$. Note that contrary to $Z$, the value of $z^\prime$ is independent of the helium abundance and reflects the mass (or number) ratios of heavy elements to hydrogen.

The BSG abundances are sensitive to the ratio of the metal fraction to the hydrogen fraction, since the opacities of these stars' atmospheres at optical wavelengths feature large contributions from bound-bound and bound-free transitions. The BSG work adopts a different set of Solar abundances from the RSG work, those of \citet{Asplund09} for oxygen and \citet{g-s98} for all the other heavy elements, which results in $z^\prime$ = 0.02.


To put the two sets of abundances on the same absolute scale, we take the ratio of the two adopted values $z^\prime$, which we call $z_{\rm B/R}$:

\begin{eqnarray}
z_{\rm B/R} \equiv z^\prime_{\rm BSG} / z^\prime_{\rm RSG} = 0.02/0.0165 = 1.212
\end{eqnarray}

\noindent Therefore, to rescale our RSG abundances to place them on the same scale as the BSG abundances, we must {\it subtract} the logarithm of this ratio, 

\begin{eqnarray}
\log( z_{\rm B/R} ) = 0.084\rm dex
\end{eqnarray}

For normal helium abundance, $n({\rm He})/n({\rm H})=0.1$, the value of $z^\prime=0.02$ adopted for this scale corresponds to $Z=0.014$. This is the same value which \citet{Asplund09} give for the proto-solar nebula.


\end{document}